\documentclass[12pt]{article}
\usepackage{amssymb,graphicx}

\makeatletter
\renewcommand\section{\@startsection {section}{1}{\z@}%
                                 {-3.5ex \@plus -1ex \@minus -.2ex}
                                   {2.3ex \@plus.2ex}%
                                   {\normalfont\large\bfseries}}
\renewcommand\subsection{\@startsection{subsection}{2}{\z@}%
                                   {-3.25ex\@plus -1ex \@minus -.2ex}%
                                     {1.5ex \@plus .2ex}%
                                     {\normalfont\bfseries}}
\renewcommand\subsubsection{\@startsection{subsubsection}{3}{\z@}%
                                   {-3.25ex\@plus -1ex \@minus -.2ex}%
                                     {1.5ex \@plus .2ex}%
                                     {\normalfont\itshape}}
\makeatother

\newcommand{\Letter}{
    \setlength{\textwidth}{7in}
    \setlength{\textheight}{9.5in}
    \hoffset=-0.75in
    \voffset=-1.15in }

\Letter

\newcommand{\non}{\nonumber \\}

\newcommand{\alp}{\alpha}     \newcommand{\bet}{\beta}
\newcommand{\gam}{\gamma}     \newcommand{\del}{\delta}
\newcommand{\eps}{\epsilon}   
      \renewcommand{\th}{\theta}

   \newcommand{\sig}{\sigma}
 
   \newcommand{\ome}{\omega}

\newcommand{\Ome}{\Omega}

\newcommand{\cG}{{\cal G}}

    \newcommand{\cN}{{\cal N}}

\newcommand{\RR}{\mathbb{R}}

\newcommand{\pa}{\partial}

\newcommand{\trp}{^{\top}}

\renewcommand{\dag}{^{\dagger}}

\newcommand{\rar}{\rightarrow}

\newcommand{\slsh}[1]{/ \!\!\! #1}

\newcommand{\hlf}{\frac{1}{2}}

\newcommand{\gsim}{ \lower .75ex \hbox{$\sim$} \llap{\raise .27ex \hbox{$>$}} }
\newcommand{\lsim}{ \lower .75ex \hbox{$\sim$} \llap{\raise .27ex \hbox{$<$}} }

\begin{document}
\thispagestyle{empty}
\begin{flushright}
\parbox[t]{2in}{
CU-TP-1172 \\
ITFA-2006-54\\
KIAS-P06066\\
hep-th/0612308}
\end{flushright}

\vspace*{0.5in}

\begin{center}
{\large \bf 
Nonextremal black holes are BPS
}

\vspace*{0.5in}
{Christopher M. Miller${}^{1}$,
Koenraad Schalm${}^{1,2}$, 
Erick J. Weinberg${}^{1,3}$
}\\[.3in]
{\em
     ${}^1$ Department of Physics, Columbia University, New York, NY
     10027 \\[.3in]
     ${}^2$ Institute for Theoretical Physics ${}^{*}$\\
     University of Amsterdam, Valckenierstraat 65, Amsterdam, 1018XE\\[.3in]
     ${}^3$ School of Physics, Korea Institute for Advanced Study\\
     207-43, Cheongnyangni2-dong, Dongdaemun-gu, Seoul 130-722, Korea
}\\[1in]
\end{center}

\begin{center}
{\bf
Abstract}
\end{center}
\noindent
Extremal charged black holes are BPS solutions.  It is commonly
thought that their nonextremal counterparts are not.  Further,
experience with BPS solutions in flat spacetime suggests that all BPS
solutions are supersymmetric; i.e. that they are invariant under some
supersymmetry charges of either the original field theory or an
appropriately extended version thereof.  Using nonextremal
Reissner-Nordstr\"om black holes as counterexamples, we show that
neither of these expectations is universally valid.  These black holes
correspond to a one-parameter family of BPS solutions.  By showing
that, subject to one very plausible assumption, no generalized
Killing spinor can be constructed for these, we show that there is no
supergravity theory for which these BPS solutions preserve a fraction
of the supersymmetry, nor is there an associated Witten-Nester
positive energy bound.

\vfill
\hrulefill\hspace*{4in}

{\footnotesize
Email addresses: \parbox[t]{5.5in}{
cmiller@phys.columbia.edu, schalm@science.uva.nl,
ejw@phys.columbia.edu.} 

${}^*$ Current address.}

\newpage

\section{Introduction}
\label{sec:introduction}

It is a well known property of supersymmetric theories that solutions
to the equations of motion that preserve a fraction of the
supersymmetries are BPS; i.e., they can be obtained by solving a set 
of equations that are first order in the fields.  It is easy to see
why this is so.  Invariance of the 
solution under a supersymmetry
implies that the
  supersymmetry variations of the fermions must vanish for the
  solution, and 
this condition gives the first-order equations.
In all known examples in flat spacetime field theory, 
the converse is also true: If
a  field theory admits BPS solutions, then it can be extended to a
supersymmetric theory, with the BPS solutions invariant under some of
the supersymmetries \cite{Sakai:2005sp,WeinbergYi}.  However, there is
no demonstration that this need be so, and one might well ask whether
there can be counterexamples.

We will address this question in this paper.  We will work in the
context of a field theory coupled to gravity.  Recall that for nongravitational
  field 
theories with BPS solutions the
energy can be written in the form
\begin{equation}
    E = \int d^3x \sum_a G_a(\phi_j, \nabla\phi_j)^2
     + \int_{|x|\rar \infty} d^2S \,  \, H(\phi_j, \nabla\phi_j)~.
\label{energyequation}
\end{equation}
Here 
the $G_a$ are linear in $\nabla\phi_j$ and the second integral,
over the sphere at spatial infinity, is completely specified by the
values of the global charges of the theory.  For fixed values of these
charges, the energy is thus bounded from below and is 
minimized by the BPS configurations obeying the first-order equations
$G_a=0$.  This is the BPS bound.
In the supersymmetric extension of the theory, the $G_a$ are
the supersymmetric variations of the fermion fields, while the surface
integral can be expressed in terms of the central charges of the
supersymmetry algebra.  

When gravity is included, static BPS solutions are naturally described
in terms of the action, rather than the energy.  One still has a
structure similar to that in Eq.~(\ref{energyequation}), but with the
energy replaced by the action, and with the crucial difference that
some of the $G_a^2$ terms enter with an overall minus sign. While it
is still true that the vanishing of all the $G_a$ produces a solution
of the full field equations of the theory, such a solution is no
longer automatically associated with the saturation of any bound.  In
this paper we will refer to solutions as being BPS if they can be
obtained from first-order equations, whether or not they saturate an
associated BPS bound.

However, there is a set of (spherically symmetric) black
hole solutions that are both BPS and associated
with a bound.  The extremal solutions whose inner and outer 
horizons
coincide are known to be supersymmetric and 
to saturate the bound imposed by the positive energy theorem.
Witten and Nester's proof of the positive energy
  theorem in fact relates these properties \cite{Witten:1981mf,Nester:1982tr}. 
In a supergravity theory one of the BPS
equations is the Killing spinor equation (the vanishing of the
variation of the gravitino), and the existence of a Killing spinor
means that the positive energy bound is saturated.  
Because the nonextremal solutions do not appear to saturate
an energy bound, it has often been thought that they are not BPS.
We will show in this paper
that this is not necessarily so.

Our interest in this problem was first aroused by trying to construct
nonextremal D3-D7-brane solutions to describe thermal gauge theories
with light flavors. Since our results are easily extended to arbitrary
static spherically symmetric $p$-branes, we will work 
instead with a simpler example, that of charged black holes in four spacetime
dimensions.  It is well known that the extremal Reissner-Nordstr\"om
black hole is BPS and supersymmetric.  We will show that despite
expectations, the nonextremal Reissner-Nordstr\"om solutions are also
BPS, and that they can be derived from a one-parameter family of
first-order equations with the parameter describing the deviation from
extremality.  The extremal solution is a limiting case.  We will see
that the underlying mechanism that allows for a family of first-order
solutions is the fact, alluded to above, that gravitational actions are
unbounded from below.

Given that all known nongravitational BPS solutions preserve a
fraction of some supersymmetry, it is natural to ask whether there is
a supersymmetric extension of the theory such that the nonextremal
black hole solutions preserve part of the supersymmetry.  In
particular, can we re-obtain the first-order BPS equations from some
supersymmetry variation of the fermion fields?  A necessary
consequence of such a preserved supersymmetry would be the existence
of a (generalized) Killing spinor corresponding to the variation
of the gravitino.  We will show that, 
subject to one mild assumption, no such spinor exists for the 
for the nonextremal solutions.  This
immediately tells us that there is no supergravity extension for which
the nonextremal BPS solutions preserve a subset of the
supersymmetries.

The remainder of this paper is organized as follows.  In Sec.~2 we
consider charged black hole solutions.  After reducing the problem to
an equivalent (1+1)-dimensional effective theory, we show that not
only the extremal, but also the nonextremal, Reissner-Nordstr\"om
black hole solutions are BPS.  We then go on to explain how the
existence of a family of BPS solutions can be understood as a
consequence of the nonpositivity of the gravitational action.  Next,
in Sec.~3, we turn to the question of supersymmetry.  We show that the
effective theory of the previous section can be extended to a
supersymmetric theory in two spacetime dimensions, and that the BPS
solutions --- both extremal and nonextremal --- preserve half of the
supersymmetry.  However, we then show that, within our
assumption, there is no lift of this supersymmetry to four
dimensions under which the nonextremal solutions are supersymmetric.
We make some concluding remarks and discuss some implications for the
saturation of energy bounds in Sec.~4.

\section{Nonextremal BPS black holes}
\label{sec:non-extremal-bps}
\setcounter{equation}{0}

The example we will use to show that nonextremal solutions
can be BPS is the $d=4$ Reissner-Nordstr\"{o}m black hole. The approach below
is easily generalized to generic static charged spherically symmetric
$p$-branes in dilaton gravity. 

The action is the Einstein-Hilbert gravitational action with a
coupling to electromagnetism; with units chosen so that $4\pi G=1$,
\begin{equation}
  \label{eq:1}
  S  = \int d^{4}x \sqrt{-g}\left[\frac{1}{4}R - \frac{1}{4}
  F_{\mu\nu}^2 \right]~.
\end{equation}
The nonextremal black hole
solution we seek is SO(3) symmetric and $t$-independent.
In a parameterization that will prove
convenient, the most general metric respecting these
symmetries can be written as
\begin{eqnarray}
  \label{eq:2}
  ds^2_{RN} = - e^{2n_4} dt^2
      +e^{-2n_4}\left[ e^{2n_3} \left(d\Ome_2^2 + e^{2n_1}du^2\right) \right]~,
\end{eqnarray}
where $d\Ome_2^2$ is the metric on the unit 2-sphere and the $n_j$ are
functions of only $u$.  Anticipating our results, we note that there
will be a horizon, at a zero of $e^{n_4}$.  We will also require that
the spacetime be asymptotically flat.  In this asymptotic region,
$e^{n_4}$ must tend to a constant which, by a rescaling of
coordinates, can be chosen to be unity.\footnote{When this metric is
generalized to $p$-branes, one must also impose the condition that
the horizon be finite and nonzero.}  For later reference, note that
these coordinates can be converted to the more usual ones by making
use of the fact that the circumference of the two-sphere is $2\pi$
times
\begin{equation}
\label{rdef}
    r(u) = e^{n_3-n_4}~.
\end{equation}

For the electromagnetic field strength we make a standard ansatz
consistent with the rotational symmetry,
\begin{eqnarray}
\label{Fansatz}
     F_{tu} &=& e^{n_1-n_3+2n_4} Q(u)~,  \cr
     F_{\th\phi} &=& 0~.
\end{eqnarray}
The electromagnetic field equation reduces to the requirement that $Q$
be constant; integrating the flux over the sphere at infinity, we see
that $4\pi Q$ is the total charge carried by the black hole.\footnote{The 
  generalization of the argument below to that of a magnetically
  charged black hole is straightforward.}

To find the solution, we will use the approach pioneered in
\cite{Gibbons:1987ps}; see also \cite{Callan:1996tv,Buchel:2001gw}.
Substituting Eqs.~(\ref{eq:2}) and (\ref{Fansatz}) into the
four-dimensional action of Eq.~(\ref{eq:1}) yields 
(after factoring out an overall $\pi$) 
an equivalent 
(1+1)-dimensional action,
\begin{equation}
    S_{\rm eff} = \int dt L_{\rm eff}
\end{equation}
where the time-independent 
effective Lagrangian is\footnote{Note that we
have not included a Gibbons-Hawking term.}
\begin{equation}
  \label{eq:3}
   L_{\rm eff} = \int du\, e^{n_3 -n_1}
   \left[2 (n_3')^2 - 2 (n_4')^2 + 2 e^{2 n_1}
         - 2 Q^2 e^{2(n_1-n_3+n_4)} \right]
    +   \int du \, \frac{d}{du}\left[ e^{n_3 -n_1}
        (2n'_4 -4 n'_3 )  \right]~.
\end{equation}
Note that $n'_1$ does not enter here, reflecting the fact that
$n_1$ can be eliminated by a coordinate transformation. 
The field equation for $n_1$ is then simply an algebraic constraint
equation.  Writing the
integrand of the Lagrangian as $e^{-n_1}T - e^{n_1} V$, this constraint 
becomes the ``zero energy condition'' $e^{-n_1}T + e^{n_1} V = 0$.

We will henceforth assume that the coordinates have been chosen so that
$n_1=n_3$.  Our effective Lagrangian then takes the form
\begin{equation}
  \label{Seff}
   L_{\rm eff} = \int du\, 
   \left[2 (n_3')^2 - 2 (n_4')^2 + 2 e^{2 n_3}
         - 2 Q^2 e^{2n_4} \right]
    +   \int du \, \frac{d}{du}\left(
       2n'_4 -4 n'_3  \right)~.
\end{equation}
The total derivative term can be ignored in deriving
the solution, but it will be important later.

\subsection{The extremal BPS solution}
\label{sec:extr-bps-solut}

This dimensionally reduced 
effective Lagrangian is easily written as a sum of
squares plus a total derivative term,
\begin{equation}
  \label{eq:7}
  L_{\rm eff} = 
     \int du\, \left[2(n_3'+e^{n_3})^2- 2(n_4'
  + Q e^{n_4})^2\right]  - 2 Z_{\rm ext}~, 
\end{equation}
where
\begin{equation}
   Z_{\rm ext} =  \int du \, \frac{d}{du}\left[
    2 e^{n_3} -2Q e^{n_4}  +2 n'_3 -  n'_4       \right]
\label{Zextdef}
\end{equation}
with the first two terms coming from completing the squares and the 
last two carried over from Eq.~(\ref{Seff}).

Requiring that each of the squares separately vanish yields a pair 
of first-order equations.  
These separate and are solved by 
\begin{eqnarray}
  \label{eq:4}
  e^{-n_3} &=& u+ c_3  = u ~, \cr 
  e^{-n_{4}} &=&  Qu + c_{4} = Qu + 1~.
\end{eqnarray}
In the second equalities we have set $c_3 =0$ by a choice of the origin
of $u$, while the boundary condition from asymptotic flatness fixes 
$c_4 = 1$.   Referring to Eq.~(\ref{rdef}), we find that 
\begin{equation}
\label{rFORextremal}
    r = {1 \over u} + Q~,
\end{equation}
so that $u = 0$ corresponds to asymptotically flat spatial infinity,
while at the horizon, $u = \infty$, we have $r(\infty) = Q$.
Converting from $u$ to $r$, we obtain the 
standard extremal Reissner-Nordstr\"{o}m solution 
\begin{eqnarray}
  \label{eq:8}
  ds^2 &=& -H\,dt^2+H^{-1}dr^2 + r^2d\Ome_2^2~, \non
 H(r) &=& \left(1- \frac{Q}{r}\right)^2~.
\end{eqnarray}

Finally, we note that substitution of our solution into
Eq. (\ref{Zextdef}) yields
\begin{equation}
   Z_{\rm ext} =  \int_0^\infty du \, \frac{d}{du} \left[
  - Q e^{n_4}\right] = Q  ~.
\end{equation}
The limits on the integral correspond to an integration over the
region outside the horizon.

\subsection{The nonextremal BPS solutions}
\label{sec:non-extremal-bps-1}

The conventional wisdom has it that 
only the extremal solution above is BPS,
and that to obtain nonextremal solutions one would have to
solve the second-order field equations.   However, with our choice of coordinates
the second-order equations separate and can each be reduced
by quadratures to a first-order equation.
This led us to surmise that one could
recognize these more general first-order equations in the Lagrangian as
well.  It takes little effort to confirm
that there is indeed a more general way to write the
$L_{\rm eff}$ in BPS form, namely
\begin{equation}
  \label{eq:11}
 L_{\rm eff} = \int du\, 
     \left[
       2\left(n_3'+ \sqrt{e^{2n_3} +\beta^2}\right)^2
  - 2\left(n_4' + \sqrt{Q^2 e^{2n_4}
      + \beta^2}\right)^2\right]  - 2 Z_{\rm non-ext}
\end{equation}
with
\begin{eqnarray}
\label{eq:6}
    Z_{\rm non-ext} &=&
\int du\, \frac{d}{du} \left[ 2 n'_3 - n'_4
         -2\sqrt{Q^2 e^{2n_4}+\beta^2} 
           +2\sqrt{e^{2n_3}+\bet^2}
        \right. \cr\cr &&\quad     \left.
         +2 \beta\, {\rm arcsinh}\, \left({\beta\over Q} e^{-n_4} \right)
          -2 \beta \, {\rm arcsinh}\, \left( \beta e^{-n_3} \right)
       \right]~.
\end{eqnarray}

As before, requiring that the two squares both vanish gives a
pair of first-order equations that separate.  Their solution is
\begin{eqnarray}
  \label{eq:14}
  e^{-n_3} &=& \frac{1}{\bet}\sinh\bet u~,  \cr
  e^{-n_{4}} &=& \frac{Q}{\beta}\sinh(\beta u + c_{4})~.
\end{eqnarray}
Here we have again absorbed the integration constant in the 
$n_3$ equation by shifting the origin of $u$.  The requirement of
asymptotic flatness gives $c_4 = {\rm arcsinh}(\beta/Q)$.
The transformation to standard coordinates is now given by
\begin{eqnarray}
  \label{eq:15}
   r(u) =  e^{n_3-n_4} = Q\frac{\sinh(\beta u+c_4)}
        {\sinh\bet u}~,
\end{eqnarray}
so that at the horizon, $u=\infty$, 
\begin{equation}
   r = r_+ = Q e^{c_4} = \sqrt{Q^2+ \beta^2} + \beta ~.
\end{equation} 

If we now define $M$ by 
\begin{equation}   
   \beta = \sqrt{M^2-Q^2 }  \, ,
\end{equation}
substitute the warp factors into the metric of Eq.~(\ref{eq:2}), and
change coordinates from $du$ to $dr$, we recover the form of
Eq.~(\ref{eq:8}), but with $H(r)$ 
taking
 the nonextremal
form
\begin{equation}
   H(r) = \left(1 - {r_+ \over r}\right)\left(1 - {r_- \over r}\right)
\end{equation}
with 
\begin{equation}
    r_\pm = M \pm \sqrt{ M^2 -  Q^2 }~.
\end{equation}
Thus, we have established that, contrary to conventional wisdom, the
nonextremal Reissner-Nordstr\"{o}m solution is BPS.

With the metric in this standard form, we recognize $M$ as the
Arnowitt-Deser-Misner (ADM)
mass, and see that $\beta$ is a measure of the nonextremality of the
black hole.  In the case of the extremal black hole, we found that the
Lagrangian of the BPS solution, which is just equal to $2Z_{\rm ext}$, was precisely
twice the ADM mass $M=Q$.  This result does not carry over to the
nonextremal case.  Instead,
\begin{eqnarray}
\label{eq:5}
   Z_{\rm non-ext} = -  \int_0^\infty du\, \frac{d}{du}\left[ 
        \sqrt{Q^2e^{2n_4}+\beta^2} -2\beta c_4\right]
     &=&  \left[ -\beta + \sqrt{ Q^2  + \beta^2} \right] \non
      &=& \left[ M -\sqrt{M^2- Q^2 }  \right]~.
\end{eqnarray}

\subsection{Families of BPS solutions}
\label{sec:arent-bps-solutions}

How is it that this system admits a family of BPS solutions?
Consider a theory involving scalar fields $\phi_i$ 
in one spatial dimension whose
static solutions are governed by an effective Lagrangian of the form
\begin{equation}
\label{LwithFF}
   L_{\rm eff} = - \int dz {\cal G}^{ij}\left[ \nabla \phi_i \nabla \phi_j
         + f_i(\phi) f_j(\phi)  \right] + 2\int dz \nabla H  ~,
\end{equation}
with ${\cal G}^{ij}$ possibly a function of $\phi$.  If the $f_i$ are such that
\begin{equation}
     f^i = {\cal G}^{ij} f_j = {\partial W \over  \partial \phi_i}
\end{equation}
for some function $W(\phi)$, then we can rewrite $L_{\rm eff}$ in the BPS 
form
\begin{equation}
\label{LwithBPS} 
   L_{\rm eff} = -\int dz {\cal G}^{ij}\left( \nabla \phi_i - f_i \right)
          \left( \nabla \phi_j - f_j \right)
                 - 2 \int dz \left(\nabla W -\nabla H \right)~.
\end{equation}

We want to know whether it is possible to rewrite this action with the 
functions $f_i$ replaced by a second set, $\tilde f_i$,
that are the derivatives of a different function, $\tilde W(\phi)$.
Clearly, this can only be done if 
\begin{equation}
\label{fCondition}
    {\cal G}^{ij} \left( f_i f_j - \tilde f_i \tilde f_j \right) = 0~.
\end{equation}
If ${\cal G}_{ij}$ has only positive eigenvalues, as would be the
case for standard scalar field theories, { 
this has the obvious
solutions $\tilde f_i = \pm f_i$.   In some
cases there
may also be isolated nontrivial solutions that lead to a distinct $
\tilde W$. The new feature in the
gravitational case, where ${\cal G}_{ij}$ is not positive definite,
is the existence of null vectors that yield continuous families of
solutions to Eq. (\ref{fCondition}).}  
If $\beta$ is such a null vector,
then Eq.~(\ref{fCondition}) is satisfied if 
\begin{equation}
     \tilde f_i \tilde f_j - f_i f_j = \pm \beta_i \beta_j~.
\end{equation} 
In a basis in which ${\cal G}^{ij}$ is diagonal, this implies that 
\begin{equation} 
     \tilde f_i = \sqrt{f_i^2 \pm \beta_i^2 }~.
\end{equation} 

In addition, the requirement that $\tilde f^i = \partial \tilde W/
\partial \phi_i$ implies that for any $i \ne j$ (and still in the 
basis that diagonalizes ${\cal G}^{ij}$)
\begin{equation} 
    {\partial \tilde f_i \over \partial \phi_j} 
        = {\partial \tilde f_j \over \partial \phi_i}   ~.
\end{equation}
If the $\beta_j$ are constants (i.e., independent of the $\phi_j$), 
it follows that 
\begin{eqnarray}
    0  &=&  {f_i  \over  \sqrt{f_i^2 \pm \beta_i^2 } } 
       \left( {\partial f_i \over \partial \phi_j } \right)
           - {f_j \over  \sqrt{f_j^2 \pm \beta_j^2} } 
      \left( {\partial f_j \over \partial \phi_i } \right) \cr\cr 
      &=& {\partial^2 W \over \partial \phi_i \, \partial \phi_j} \left[ 
            \left( 1 \pm {\beta_i^2 \over f_i^2} \right)^{-1/2} 
          -\left( 1 \pm {\beta_j^2 \over f_j^2} \right)^{-1/2} \right]~.    
\end{eqnarray}
The quantity in brackets in the last line only vanishes if $\beta_j^2/f_j^2$
is the same for all values of $j$. This would imply that $f$, like $\beta$,
would be a null vector and so would not contribute to the action at all.  The only
alternative, then, is that the mixed second derivatives of $W$ all vanish; i.e.,
that each of the $f_i$ be a function of only the corresponding $\phi_i$.
Indeed, a quick glance at Eq.~(\ref{eq:11}) shows this is precisely the case for our
Reissner-Nordstr\"{o}m black holes.

\subsection{Extremality and the BPS bound} 
\label{sec:extr-bps-bound}

In the nongravitational examples, the BPS solutions saturate a lower
bound on the energy.  Although there is no such BPS bound for the
Lagrangian in our gravitational example, these systems do 
respect a 
physically inspired bound.  To avoid a naked
singularity, charged black holes must obey the positive energy
theorem, which requires that their ADM energy be greater than or equal
to their charge; the work of Witten and Nester showed that this bound
is the analogue of the BPS bound for gravitational systems
\cite{Witten:1981mf,Nester:1982tr,Gibbons:1982jg}.
This bound is, by definition, saturated by extremal black holes with
ADM energy $M$ equal to the charge $Q$. 

In our BPS solutions the Lagrangian was equal to twice a quantity $Z$, whose relation
to a central charge we will discuss in the next section. For the present, we
just restate that while in the extremal case we have $Z_{\rm ext} = M$, the 
corresponding equality does not hold for the nonextremal solutions, where
$Z_{\rm non-ext} < M$.

\section{Supersymmetry and the BPS bound}
\label{sec:supersymm-bps-bound}
\setcounter{equation}{0}

The intimate connection between BPS first-order solutions and
solutions preserving a fraction of supersymmetries raises the question
of whether the full one-parameter family of nonextremal solutions
preserves some supersymmetry --- whose form might depend on the value
of the nonextremality parameter $\beta$ ---
or whether the extremal solution is 
special in that regard. We will show in this section that
the full one-parameter family of black hole solutions in the previous
section can preserve half the supersymmetries in an appropriate
(nongravitational) supersymmetric extension.
The Lagrangian of Eq. (\ref{Seff}) defines a (1+1)-dimensional linear
sigma model with a complicated potential, but all first-order 
solutions 
in an $\cN=2$ extension of such a theory preserve half the
supersymmetries. 

This need not imply, however, that the nonextremal solutions preserve
some of the supersymmetries of a supersymmetric extension of the
original $d=4$ Einstein-Maxwell gravity. There could be an obstruction
to lifting the preserved supersymmetries in the lower-dimensional
effective theory back to four dimensions.  In fact, we know beforehand
that this extension cannot be trivial $\cN=2$, $d=4$ supergravity. The
BPS states of that theory have long since been known and are limited
to the extremal ones \cite{Gibbons:1982fy} (see also
\cite{Caldarelli:2003pb,Cacciatori:2004rt}). 
Any putative supergravity for which nonextremal
solutions preserve a subset of the symmetries must therefore be a
novel one.  

Rather than searching directly for this novel supergravity, we address
the issue by looking for a generalized Killing spinor.  Any preserved
supersymmetry in a supergravity theory must yield such a spinor, from
the variation of the gravitino.  (The converse, however, need not be
true; the existence of a Killing spinor does not imply supersymmetry.
We will return to this point below.)  We will show, modulo a mild
assumption that we explain below, that only the extremal solution
allows such a Killing spinor.  Hence, the nonextremal solutions
cannot be supersymmetric.

\subsection{Preserved supersymmetries in the lower-dimensional
  effective theory}
\label{sec:pres-supersymm-lower}

The effective Lagrangian of Eq.~(\ref{Seff}) describes dynamics in one spatial
dimension. The extremal and nonextremal solutions to the field
equations have a nontrivial profile only along this dimension, and
they are therefore equivalent to static solitons in a (1+1)-dimensional
field theory.  The appropriate effective supersymmetric extension to
consider is therefore an $\cN=2$, (1+1)-dimensional theory. This theory
is the reduction of 
a 
minimally supersymmetric $\cN=1$ 
theory 
in 3+1
dimensions. Its dynamics are well known and in component
language the most general action
is\footnote{Our conventions are such that
$\bar{\phi}^{\bar{a}}=(\phi^a)^{\ast}$ is the complex conjugate of the
scalar field $\phi^a$; $\bar{\psi}^{\bar{a}} = (\psi^a)^{\dag} i
\gam^0$ is the Dirac conjugate, and $\hat{\psi}^a =
(\psi^a)^{\trp}i\gam^0 = (\psi^a)^{\trp}C$ the Majorana conjugate of
the two-component complex spinor $\psi^a$; and
$\{\gam^{\mu},\gam^{\nu}\}=2\eta^{\mu\nu}$ with $\gam^0$
anti-Hermitian and $\gam^1$ Hermitian.
Recall that in 1+1 dimensions all supersymmetry multiplets are dual to
  the chiral multiplet.}
\begin{eqnarray}
  \label{eq:28}
  S^{\cN=2}_{1+1} &=& \int dx \, dt \Big[
  -\cG_{\bar{a}b}\left[\nabla_{\mu}\bar{\phi}^{\bar{a}}\nabla^{\mu}{\phi}^{{b}}
  +  \bar{\psi}^{\bar{a}}\slsh{\nabla}\psi^b -
  \bar{F}^{\bar{a}}F^{b}\right] + \bar{F}^{\bar{a}}\pa_{\bar{a}}\bar{W} +
  F^b\pa_{{b}}W
\Big.
\non
&& \hspace{.3in}~~~~\left.
-\hlf\left(\hat{\psi}^a\psi^b \pa_{a}\pa_b
  W +
(\bar{\psi})^{\bar{a}}(\psi^*)^{\bar{b}}\pa_{\bar{a}}\pa_{\bar{b}}\bar{W}\right)
\right]~.
\end{eqnarray}
Here $W(\phi^a)$ is the holomorphic superpotential of the fields and
the metric $\cG_{a\bar{b}}= \pa_a\pa_{\bar{b}}K(\phi,\bar{\phi})$ is
the mixed second derivative of the real Kahler potential; to avoid 
confusion between derivatives with respect to fields and spacetime
derivatives, we have denoted the latter by $\nabla_\mu$.
By construction the action is invariant under the supersymmetry
transformations
\begin{eqnarray}
  \label{eq:29}
  \del \phi^a = \bar{\eps}\psi^a~,~~ \del\psi^a = \slsh{\nabla}\phi^a\eps
  + F^a{\eps}^*~,~~\del F^a = \hat{\eps}\slsh{\nabla}\psi^a ~.
\end{eqnarray}
Because $\eps$ is a two-component complex spinor, there are four
supercharges.  

As is well known, integrating out the the auxiliary scalar field $F^a
= -\cG^{a{\bar{b}}}\pa_{\bar{b}}\bar{W}$ yields a scalar field potential
$\cG^{a{\bar{b}}} \pa_a W \pa_{\bar{b}}\bar{W}$ that, for real fields
$\Phi = \hlf (\phi+\bar{\phi})$, is precisely of the form that appears
in Eq.~(\ref{LwithFF}).  For solutions of the corresponding BPS
equation, $\nabla_1\Phi^a + \cG^{a\bar{b}}\pa_{\bar{b}}\bar{W}(\Phi) =0$,
the effect of a supersymmetry transformation on the fermion,
\begin{eqnarray}
  \del \psi^a &=& \gam^1\nabla_1\Phi^a\eps -
    \cG^{a\bar{b}}\pa_{\bar{b}}\bar{W}(\Phi){\eps}^* \\ \nonumber
&=& \nabla_1 \Phi^a \frac{(1+\gam_1)}{2}\eps  -
    \pa^a\bar{W}(\Phi)\frac{(1+\gam_1)}{2}\eps^* - \nabla_1 \Phi^a
\frac{(1-\gam_1)}{2}\eps  -
    \pa^a\bar{W}(\Phi)\frac{(1-\gam_1)}{2}\eps^*  ~,
\end{eqnarray}
vanishes if $\hlf{(1+\gam_1)}\,{\rm Im}\,(\eps) =
\hlf{(1-\gam_1)}\,{\rm Re}\,(\eps) =0$.  Thus, the BPS solutions
preserve half of the supersymmetry.  Furthermore, it is easy to
verify, by steps completely analogous to those in
Eqs.~(\ref{LwithFF})-(\ref{LwithBPS}), that for the BPS solutions the one-dimensional
effective Lagrangian is just equal to twice the central charge,
\begin{equation} 
    Z_{\rm SUSY} =  \int_{x_{\rm min}}^{x_{\rm max}} \nabla W 
          =  W(x_{\rm max})-  W(x_{\rm min})~.
\end{equation}

Applying this to the nonextremal black hole example of the previous section,
we see that $W$ differs from the quantity appearing in the integrand
of $Z_{\rm non-ext}$ in Eq.~(\ref{eq:6}) 
by the surface terms inherited from Eq.~(\ref{Seff}); i.e., 
\begin{equation}
     W = 
         -2\sqrt{Q^2 e^{2n_4}+\beta^2} 
           +2\sqrt{e^{2n_3}+\bet^2}
         +2 \beta\, {\rm arcsinh}\, \left({\beta\over Q} e^{-n_4} \right)
          -2 \beta \, {\rm arcsinh}\, \left( \beta e^{-n_3} \right) ~.
\end{equation}
Substituting the BPS solution into this yields
\begin{equation}
     W(u) =  -2 \beta \frac{\cosh(\beta u + c_4)}{\sinh
       (\beta u + c_4)} + 2\beta \frac{\cosh(\beta u)}{\sinh (\beta
       u)}+2\beta^2(u+c_4)-2\beta^2 u ~,
\end{equation}
which diverges as $u \rightarrow 0$ (i.e., as $r \rightarrow \infty$).
Hence, $Z_{\rm SUSY} = W(\infty) - W(0)$ will be infinite.  In
Sec.~\ref{sec:non-extremal-bps-1} this divergence was cancelled by the
inclusion in $Z_{\rm ext}$ and $Z_{\rm non-ext}$ of the additional
surface term appearing in Eq.~(\ref{eq:6}).  This can be interpreted as
the subtraction of the (also
infinite) central charge of Minkowski space.

\medskip

\subsection{Is there a Killing spinor?}
\label{sec:supersymm-embedd}

If nonextremal black holes are also
supersymmetric in some 
  four-dimensional theory,
there must be Killing spinors corresponding to the preserved supersymmetries
for which the variations of the gravitinos $\psi_{\mu j}$ and the spin-1/2
fields $\chi_J$ vanish.   Thus, there must be spinors $\epsilon_j$ such that 
\begin{eqnarray}
  \label{eq:58}
  0 &=&  \delta \psi_{\mu i} = (\widehat{\nabla}_{\mu})_i{}^j \eps_j ~,\non
  0 &=& \delta \chi_I     = (R^{j}_I)\eps_j ~.
\end{eqnarray}
Here the generalized covariant derivative
$(\widehat{\nabla}_{\rho})^{ij}\eps_j$ 
and the functional $R^j_I$ can in principle depend on any of the bosonic fields of
the
theory. Naturally the generalized covariant derivative must contain
the standard spin connection; its
most general form is therefore
\begin{eqnarray}
  \label{eq:53}
  \widehat{\nabla}_{\rho}^{ij}\eps_j
  &=& \left({\nabla}_{\rho}\del ^{ij} + Y_{\rho}^{ij}\right)\eps_j \non
&=&
\left[\pa_{\rho}\delta^{\alp}_{\bet}\delta^{ij}
+\ome_{\rho}^{ab}\frac{1}{4}(\gamma_{ab})_{\beta}^{\alp}\delta^{ij}
+ (Y_{\rho}^{ij})_{\bet}^{\alp}\right] \eps_j^\bet ~.
\end{eqnarray}
In the second line we have explicitly written out the spinor indices
$\alp,~\bet$.  All (possibly nonlinear) dependence on the
nongravitational bosonic fields of the theory is contained in  
$(Y_{\rho}^{ij})^{\alp}_{\bet}$ and $R^j_I$.
This dependence must be such that the
commutators of the supersymmetry variations are consistent with the
supersymmetry algebra.  We will assume that this is possible only if
these quantities vanish whenever the nongravitational fields all
vanish.  For our solutions, this means that $Y_\mu$ must vanish if
$F_{\mu\nu}=0$.

If such a Killing spinor existed, it would lead to a generalized BPS
bound through the positive energy theorem.  However, the relation of
this bound to supersymmetry is not manifest.  Even though a Killing
spinor which implies a saturation of the positive energy bound may
exist, it is not evident that such Killing spinor arises as the
gravitino variation of a known supergravity theory. In fact recent
research has shown that quite a number of spacetimes exist which are
BPS, saturate a positive energy bound, have a Killing spinor, but do
not preserve any of the supersymmetries of any known
supergravity theory
\cite{Freedman:2003ax,Celi:2004st,Zagermann:2004ac,Skenderis:2006jq,Skenderis:2006fb}.
Thus, even though demonstrating the existence of a Killing spinor
would establish that the nonextremal solutions saturate a modified
energy bound, it would only be a first step toward showing that they
preserve some supersymmetry.

On the other hand, if it can be shown that there is no Killing spinor,
then there is no preserved supersymmetry and, presumably, no saturated bound
associated with these solutions.   We will now proceed to show that this
is the case.  We will do this by showing that there is no choice for
$Y_\mu$ that (1)
when evaluated on the nonextremal solution
admits a nonzero spinor satisfying 
\begin{eqnarray}
  \label{eq:62}
  (\nabla_{\mu}+Y_{\mu})\eps  = 0~.
\end{eqnarray}
and (2) vanishes when all of the nongravitational fields are set equal 
to zero.

We start by substituting 
the nonextremal Reissner-Nordstr\"om 
metric into the Killing
equation to determine what form $Y_\mu$ can take.
In doing this, it will be helpful to perform some conformal
transformations on the metric.  We first recall that 
under a conformal rescaling $\tilde{g}_{\mu\nu} = e^{2A}g_{\mu\nu}$ ,
the covariant derivative $\nabla_\mu \epsilon$ of a spinor is
transformed to 
\begin{eqnarray}
  \label{eq:9}
  \tilde \nabla_{\mu} \eps = \left[\nabla_{\mu} +
  \frac{1}{2}\gam_{\mu}^{~\nu} \,\nabla_{\nu} A\right]\eps  ~,
\end{eqnarray}
where $\gamma^{\mu \nu} = {1 \over 2}[\gamma^\mu,\gamma^\nu]$.  

Now let us define the metrics 
\begin{eqnarray}
\label{eq:10}
    ds^2_{(4)} &=& e^{-2n_4} ds^2 = -dt^2 + e^{2n_3-4n_4}
        \left(d\Ome_2^2 + e^{2n_3}du^2\right)~,
     \cr \cr
    ds^2_{(3)} &=& -dt^2 + \left(d\Ome_2^2 + e^{2n_3}du^2\right)~,
    \cr \cr
    ds^2_{(1)}&=& -dt^2 + d\Ome_2^2  + du^2~.
\end{eqnarray}
We will
use superscripts $(4)$ and $(3)$ to
denote the
covariant derivatives and gamma matrices corresponding to $ds^2_{(4)}$
and $ds^2_{(3)}$; the gamma matrices corresponding to $ds^2_{(1)}$ will
be indicated by hats.

The first of these metrics, $ds_{(4)}^2$, is just an overall conformal
rescaling of our metric $ds^2$.  Hence, using Eq.~(\ref{eq:9}), we
obtain
\begin{eqnarray}
  \label{eq:12}
  \nabla_{\mu} \eps &=& \left[\nabla^{(4)}_{\mu} + \hlf
  \gam^{(4)\nu}_{\mu} \, \nabla_{\nu}n_4 \right]\eps \non
&=& \left[\nabla^{(4)}_{\mu} + \hlf
  \gam^{(4)u}_{\mu} \, \nabla_{u}n_4 \right]\eps~.
\end{eqnarray}
In the last step we have used the fact that $n_4$ only depends
on the coordinate $u$.

We next note that
the metric $ds^2_{(4)}$ is a direct product between $\RR$
(time) and a three-dimensional space spanned by $u$, $\theta$, and $\phi$.
We can therefore decompose the spinor
covariant derivative into two equations and perform a second conformal
rescaling, just on the directions $m=u,\th,\phi$. 
As a result the 
  connections and gamma matrices are those for the metric
  $ds_{(3)}^2$.
Again using Eq.~(\ref{eq:9}), we find that
  \begin{eqnarray}
    \label{eq:13}
     \begin{array}{rcrcl}
\nabla_t \eps &=& \left\{ \nabla_{t}^{(4)}  + \hlf
  \gam^{(4)u}_{\mu} \,\nabla_{u}n_4  \right\}\eps
    &=& \left\{{\displaystyle\frac{\partial}{\partial
    t}} + \hlf
  \gam^{(3)u}_t e^{-n_3+2n_4} \,\nabla_{u}n_4\right\}\eps~, \\ \\
\nabla_m \eps &=&
  \left\{\nabla_{m}^{(4)} + \hlf
  \gam^{(4)u}_{\mu} \,\nabla_{u}n_4 \right\}\eps
    &=& \left\{\nabla_m^{(3)} + \hlf
  \gam^{(3)u}_{m} [\nabla_u(n_3-2n_4)] + \hlf
  \gam^{(3)u}_m \,\nabla_u n_4 \right\}\eps ~.
\end{array}
  \end{eqnarray}
Implicit here is a simultaneous decomposition of the
(3+1)-dimensional spinor and Dirac matrices into one- and three-dimensional
ones; we will make this explicit shortly.

By construction the spatial part of $ds^2_{(3)}$ 
is again a direct product and so we can, by a similar procedure,
transform to $ds^2_{(1)}$ via a conformal transformation on the 
one-dimensional radial part of the metric, leading to 
\begin{eqnarray}
  \label{eq:16}
      \begin{array}{rcrcl}
\nabla_t\eps &=&
      \left\{{\displaystyle\frac{\partial}{\partial
    t}} + \hlf
  \gam^{(3)u}_{t} \,  e^{-n_3+2n_4} \,\nabla_{u}n_4\right\}\eps
&=&
\left\{{\displaystyle \frac{\pa}{\pa t}} 
+ \hlf \hat{\gam}_{t}^{~u} \, e^{2(n_4-n_3)} \nabla_u
  n_4 \right\}\eps~,
\\  \\
\nabla_{\th} \eps &=&  \left\{\nabla_{\th}^{(3)} + \hlf
  \gam^{(3)u}_{\th} \, [\nabla_u(n_3-n_4)]  \right\}\eps  
&=& \left\{ \nabla_{\th}^{(S^2)} + \hlf \hat{\gam}_{\th}^{~u}
  \,  e^{-n_3}[\nabla_u(n_3-n_4)]\right\}\eps~,
\\  \\
\nabla_{\phi}\eps &=& \left\{\nabla_{\phi}^{(3)} + \hlf
  \gam^{(3)u}_{\phi} \, [\nabla_u(n_3-n_4)]  \right\}\eps 
&=& \left\{ \nabla_{\phi}^{(S^2)} + \hlf \hat{\gam}_{\phi}^{~u}
  \, e^{-n_3}[\nabla_u(n_3-n_4)]\right\}\eps~,
\\  \\
\nabla_u\eps &=&
\left\{\nabla_u^{(3)} + \hlf
  \gam^{(3)u}_{u} \, [\nabla_u(n_3-n_4)]  \right\}\eps   
&=& {\displaystyle \frac{\pa}{\partial u} }\eps~.
\end{array}
\end{eqnarray}
This makes the full $u$-dependence of the covariant derivative explicit.

Inserting now the appropriately rescaled generalized connection
$Y_{\mu}=e_{\mu}^a\hat{Y}_a$ with $e_{\mu}^a$ equal to
$\delta_{\mu}^a$ times the appropriate conformal factor as described
in Eq.~(\ref{eq:10}), we find that the conditions for the existence of
a generalized Killing spinor are
\begin{eqnarray}
  \label{eq:63}
   0 &=&  \left\{{\displaystyle \frac{\pa}{\pa t}} 
+ \hlf \hat{\gam}_{t}\hat{\gam}^{u} e^{2(n_4-n_3)} \nabla_u
  n_4  + e^{n_4} \hat{Y}_t\right\}\eps ~, \non
  0 &=&\left\{\nabla_{\th}^{(S^2)} + \hlf \hat{\gam}_{\th}\hat{\gam}^{u}
  e^{-n_3}[\nabla_u(n_3-n_4)]+
   e^{n_3-n_4} \hat{Y}_{\th}\right\}\eps
   ~,\non
  0 &=& \left\{\nabla_{\phi}^{(S^2)} + \hlf \hat{\gam}_{\phi}\hat{\gam}^{u}
  e^{-n_3}[\nabla_u(n_3-n_4)]+
   e^{n_3-n_4} \hat{Y}_{\phi}\right\}\eps
   ~,\non
 0 &=& \left\{\frac{\partial}{\pa u} + e^{2n_3-n_4} \hat{Y}_u\right\}\eps ~.
\end{eqnarray}
The decomposition of the spacetimes $ds_{(i)}^2$ into
direct products, which we used to implement the conformal rescalings,
implies that the (3+1)-dimensional spinor $\eps$ can be written in the form
\begin{eqnarray}
  \label{eq:17}
  \eps(t,\th,\phi,u) = \eta(t,u) \otimes \zeta(\th,\phi) ~.
\end{eqnarray}
The time-independence of all the Killing vectors of the metric
of Eq.~(\ref{eq:2}) demands that the Killing spinor $\eps$ must be
time-independent. Then the first equation of (\ref{eq:63}) implies 
that $\hat{Y}_t$ must be such that on 
the Reissner-Nordstr\"om background
\begin{eqnarray}
  \label{eq:64}
   e^{n_4-2n_3}\hat{\gam}^{u}\nabla_u n_4 \, \epsilon 
     =  -\hat{\gam}^t\hat{Y}_t  \,\epsilon  
\end{eqnarray}
where here (and in similar cases below) there is no sum 
over $u$ or $t$.
Substituting this background expression for $\nabla_u n_4$ into 
the second and third equations of (\ref{eq:63}),
we obtain
\begin{eqnarray}
  \label{eq:65}
  \left[\nabla_{\th}^{(S^2)}
    +\hlf\hat{\gam}_{\th}\hat{\gam}_{u} e^{-n_3}\nabla_u n_3  +
    e^{n_3-n_4}\hat{\gamma}_{\theta}\left(\hat{\gam}^t\hat{Y}_t+
      \hat{\gam}^{\theta}\hat{Y}_\theta \right)\right]\eps&=&0~,\non
  \left[\nabla_{\phi}^{(S^2)}
    +\hlf\hat{\gam}_{\phi}\hat{\gam}_{u} e^{-n_3}\nabla_u n_3  +
    e^{n_3-n_4}\hat{\gamma}_{\phi}\left(\hat{\gam}^t\hat{Y}_t+
      \hat{\gam}^{\phi}\hat{Y}_\phi \right)\right]\eps&=&0~.
\end{eqnarray}
Compatibility with the decomposition of the Killing spinor of
Eq.~(\ref{eq:17}) demands that we find a spinor $\zeta$ on the
unit two-sphere that obeys
\begin{eqnarray}
  \label{eq:18}
  \nabla_{\th}^{(S^2)}\zeta = \sigma_3^{(S^2)}\sigma^{(S^2)}_{\th} A\zeta ~,
    ~~\nabla_{\phi}^{(S^2)}\zeta = \sigma_3^{(S^2)}\sigma^{(S^2)}_{\phi} A\zeta ~.
\end{eqnarray}
with $A$ independent of $\th,\phi$. The choice to include an additional
factor of $\sigma_3^{(S^2)}$ is made to ensure that $A$ is real and
that there exists a
Killing spinor for the extremal case; see below.
Acting twice with Eq.~(\ref{eq:18}) gives the integrability
  constraint
\begin{eqnarray}
    \frac{1}{4} R_{\th\phi}^{(S^2)}{}^{\alpha\beta} \sig^{(S^2)}_{\alpha \beta} \zeta
   =   \left(\nabla_{\th}^{(S^2)} \nabla_{\phi}^{(S^2)} 
   -\nabla_{\phi}^{(S^2)} \nabla_{\th}^{(S^2)}\right)\zeta
    =  2\sigma^{(S^2)}_{\th\phi} A^2 \zeta \, .
\end{eqnarray}
As $S^2$ is the unit two-sphere, it follows that $A=\pm
\frac{1}{2}$. If we embed the
$S^2$
Dirac matrices $\sigma_{\th}$,~$\sigma_{\phi}$ into the four-dimensional ones via
\begin{eqnarray}
  \label{eq:1b}
  \hat{\gamma}^t &=& i\sigma_2 \otimes \sigma_3 ~,\non
  \hat{\gamma}^\theta &=& \sigma_3 \otimes \sigma_{\theta} = 1 \otimes
  \sigma_1 ~, \non 
  \hat{\gamma}^\phi &=& \sigma_3 \otimes \sigma_{\phi} = 1 \otimes
  \sigma_2 ~,\non 
  \hat{\gamma}^u &=& \sigma_3 \otimes \sigma_3~, 
\end{eqnarray}
Eq.~(\ref{eq:65}) reduces to the requirement that
\begin{eqnarray}
  \label{eq:20}
  \left[\hlf\hat{\gam}_{u} \left(e^{-n_3}\nabla_u n_3 \pm 1\right)
  + e^{n_3-n_4}\left(\hat{\gam}^t\hat{Y}_t+
      \hat{\gam}^{\theta}\hat{Y}_\theta \right)\right]\eps&=&0~,\non
  \left[\hlf\hat{\gam}_{u} \left(e^{-n_3}\nabla_u n_3 \pm 1\right)
   + e^{n_3-n_4}\left(\hat{\gam}^t\hat{Y}_t+
      \hat{\gam}^{\phi}\hat{Y}_\phi \right)\right]\eps&=&0~.
\end{eqnarray}

We next decompose $Y_{\mu}$ as 
\begin{equation}
 Y_{\mu}=\gam_{\mu}Y+ C_{\rho\sig}\gam^{\rho\sig}\gam_{\mu} \, ,
\end{equation}
where $Y$, 
but not $C_{\rho\sig}$, is a
matrix in spinor space. Because
Lorentz transformations must act in the standard way, we can do so
without loss of generality.\footnote{The need to separate
 out $C_{\rho\sig}$ is due to the $d=4$ identity
 $\gam^{\mu}\gam^{\rho\sig}\gam_{\mu}=0$. To prove that $Y_{\mu}$ can
 always be written as
 $Y_{\mu}=\gam_{\mu}Y+\gam^{\rho\sig}C_{\rho\sig}\gam_{\mu}$, note that
 $Y_{\mu}-\gam^{\rho\sig}C_{\rho\sig}\gam_{\mu} 
     =\sum_{n=0}^{d}Z_{\alp_1\ldots\alp_n\mu}\gam^{\alp_1\ldots\alp_n}$ 
 can be uniquely reconstructed from
 $Y=\sum_{n=0}^{d}X_{\alp_1\ldots\alp_n}\gam^{\alp_1\ldots\alp_n}$
 through $Z_{\alp_1\ldots\alp_n\mu} =\frac{1}{n!}{\rm
 Tr}(\gam_{\mu}Y\gam_{\alp_1\ldots\alp_n})$. Using that
 $\gam_{\alp_1\ldots\alp_n} \gam_{\mu}= \gam_{\alp_1\ldots\alp_n\mu}+
 \frac{1}{(n-1)!}\gam_{[\alp_1\ldots\alp_{n-1}}\eta_{\alp_n]\mu}$, one
 finds that $Z_{\alp_1\ldots\alp_n\mu} = X_{\alp_1\ldots\alp_n\mu} +
 \frac{1}{n!(n-1)}X_{[\alp_1\ldots\alp_{n-1}}\eta_{\alp_n]\mu}$. } 
The spherical symmetry of the background implies that the only nonzero
components of $C_{ab}$ are $C_{tu}$
and $C_{\theta\phi}$. As a consequence of the Dirac matrix identities  
\begin{eqnarray}
C_{tu}\left(\gamma^t\gamma^{tu}\gamma_t +
\gamma^{\theta}\gamma^{tu}\gamma_{\theta}\right)= 0~,\non
C_{\theta\phi}\left(\gamma^t\gamma^{\theta\phi}\gamma_t +
\gamma^{\theta}\gamma^{\theta\phi}\gamma_{\theta}\right)= 0~,
\end{eqnarray}
the $C_{ab}$ parts vanish identically in Eq.~(\ref{eq:20}), 
leaving the single equation
\begin{equation}
\label{eq:3c}  
\left[\hlf\hat{\gam}_{u} \left(e^{-n_3}\nabla_u n_3 \pm 1\right)
  + e^{n_3-n_4}2Y\right]\eps =0.
\end{equation}

For the extremal solution, this is solved by taking the upper sign in the
first term and setting $Y=0$.  It is then straightforward to show that the 
remaining equations can be solved, and hence that there is a Killing spinor,
if $C_{\mu\nu} = F_{\mu\nu}$. 

 For the nonextremal solutions, we would need a $Y$ constructed out of
the background bosonic fields that was such that
\begin{eqnarray}
4Y\eps &=& - e^{-n_3+n_4}\gamma^u(e^{-n_3}\nabla_u n_3\mp 1)\eps~.
\label{nonextYeq}
\end{eqnarray}
We could proceed by evaluating the quantity on the right-hand side and 
then trying to express it in terms of the background fields.   However, there is
no need to do so.   We are assuming that any choice of $Y_\mu$ that gives a
gravitino variation consistent with the superalgebra must 
vanish when the nongravitational fields are set equal to zero.  In the present
case, the only nongravitational field is $F_{\mu\nu}$, which clearly vanishes
when $Q=0$.  However, substituting the $Q=0$ nonextremal solution  yields
\begin{equation}
e^{-n_3}\nabla n_3 \mp 1 = - \hlf \sqrt{(1-M/r)+
    \frac{1}{(1-M/r)}+2} \, \mp 1 \ne 0
\end{equation}
Except in the trivial $M=0$ case, Eq.~(\ref{nonextYeq}) then requires a nonzero $Y$,
in contradiction to our assumption.  Hence, the nonextremal solutions do not
admit Killing spinors.

\section{Discussion and Conclusion}
\label{sec:disc-concl}

We have shown that static nonextremal black hole solutions are BPS;
i.e., they are solutions to first-order equations obtained by
writing the Lagrangian in terms of a sum of squares.  As we explained
in the introduction, all known BPS solutions in flat spacetime field
theory preserve a fraction of supersymmetries in some extension of the
theory.  However, we have seen that this empirical result cannot
extend to our gravitational theory

For the dimensionally reduced (1+1)-dimensional theory it is 
a straightforward matter to solve the 
Killing spinor equations and find a preserved supersymmetry.  The difficulty 
is with the lift to (3+1) dimensions.   If we make the plausible assumption
that consistency with the supersymmetry algebra requires that any additional
terms in the generalized connection must vanish when the nongravitational fields
are all zero, we then find that the (3+1) nonextremal solutions admit
no Killing spinors.  Hence, there cannot be a supergravity extension of the theory where
the nonextremal BPS solutions preserve a fraction of the supersymmetries.

We can thus distinguish four types of gravitational
solutions: (i) first-order solutions that preserve a fraction of
supersymmetries in a known supergravity theory, (ii) first-order
solutions that saturate a bound following from the existence of a
generalized Killing spinor, but that do not correspond to a
known gravitino
variation, (iii) first-order solutions that do not appear to saturate a bound,
and (iv) standard second-order solutions. In this
note we have shown 
that nonextremal black branes belong to category (iii) and
why they do so.

Finally, we wish to point out that our results share a connection with
Sen's recent construction of an entropy function for extremal black holes 
\cite{Sen:2005wa,Dabholkar:2006tb,Cardoso:2006xz}. For extremal black
holes with near-horizon $AdS_2\times S^2$ geometry this entropy
function is the Legendre transform with respect to the electric charges of the
Lagrangian density integrated over the $S^2$. This two-sphere is conformal to
the one over which we reduced to construct our
(1+1)-dimensional effective action. Hence, the (1+1)-dimensional
Lagrangian density for which both extremal and nonextremal black
holes are BPS solutions is related to Sen's entropy function. It would be
interesting to investigate whether this entropy function can be used for
nonextremal black holes as well. {A naive extension of Sen's entropy
function to nonextremal black 
holes does not work, as it relies crucially on the $AdS_2\times S_2$
near horizon geometry of extremal black holes rather than the BPS
structure \cite{Astefanesei:2006sy}.}

\bigskip

{\bf Added note:}
After this work was completed, we learned of Ref.~\cite{Lu:2003iv},
where it was noted that some nonextremal anti-de Sitter black holes
can be obtained from first-order equations that integrate to a
superpotential. The methods we use in
Sec.~3 are readily generalized to this case, and show that these
nonextremal black holes are not supersymmetric.

\bigskip

{\bf Acknowledgements} We are grateful to David Tong for
correspondence and to Carlos Nunez and Kostas Skenderis for
discussions, {
and to Ed Witten for
pointing out an error in an earlier version of this manuscript.}  
CM thanks the Korea Institute for Advanced Study, and KS
thanks the Perimeter Institute for their generous hospitality.  This
research has been supported in part by DOE grant DE-FG-02-92ER40699
and a VIDI Research Grant from the Netherlands Organisation for
Scientific Research (NWO) (KS). The authors gratefully acknowledge
support from the Ohrstrom Foundation.

\end{document}